\documentclass[a4paper]{jpconf}
\usepackage{graphicx}
\usepackage{latexsym}
\usepackage{amsfonts}
\usepackage{color}
\usepackage{colortbl}
\usepackage{amssymb,amsmath,enumerate}
\usepackage{amsthm} 
\usepackage{bm}
\usepackage{indentfirst}
\usepackage{mathrsfs}
\usepackage[T1]{fontenc}
\usepackage{psfrag}

\newtheorem{lemma}{Lemma}
\newtheorem{proposition}{Proposition}

\newcommand{\Real}{\mathbb{R}}

\begin{document}

\title{On the behaviour of non-radial null geodesics in self-similar Tolman-Bondi collapse}

\author{N\'estor Ortiz$^{1,2}$, Olivier Sarbach$^{1,2}$, and Thomas Zannias$^{1,3}$}

\address{$^1$ Instituto de F\'{\i}sica y Matem\'aticas,
Universidad Michoacana de San Nicol\'as de Hidalgo\\
Edificio C-3, Ciudad Universitaria, 58040 Morelia, Michoac\'an, M\'exico.}

\address{$^2$ Perimeter Institute for Theoretical Physics, 31 Caroline St., Waterloo, ON, N2L 2Y5, Canada.}

\address{$^3$ Department of Physics, Queen's University, Kingston, Ontario K7L 3N6, Canada.}

\ead{nortiz@perimeterinstitute.ca, sarbach@ifm.umich.mx, zannias@ifm.umich.mx}

\begin{abstract}
Motivated by recent work on the structure of the singularity in inhomogeneous Tolman-Bondi collapse models, we investigate the behaviour of null geodesics in the particular case where the collapse is self-similar. The presence of the homothetic Killing vector field implies that the geodesic equation can be described by an integrable Hamiltonian system, and exploiting this fact we provide a full qualitative picture for its phase flow.
\end{abstract}

\section{Introduction}

Our late colleague and friend Victor Villanueva had broad interests in theoretical physics, and although he was considered a particle physicist, he also made important contributions to the field of general relativity. In particular, Victor worked on the geodesic deviation equations for relativistic spinning particles and possible applications to the detection of gravitational radiation~\cite{vV03,vV07,vV94}, a topic that is tightly related to upcoming experiments regarding the detection of gravitational waves through the use of pulsar timing arrays~\cite{IPTA}. More recently, Victor also worked on the Hamiltonian formulation for higher-dimensional black holes~\cite{vV13}. We dedicate the present work to him. We are confident he would have shared our curiosity in the problem analyzed in this article.

It is well known that a Tolman-Bondi spacetime, describing the collapse of  a spherical dust cloud, admits shell-focusing singularities, a portion of which may be null and visible to local observers~\cite{dElS79,dC84,rN86,pJiD93,Joshi-Book,bNfM01,nOoS11}. Furthermore, for suitable initial data, part of this null singularity may also be visible to observers in the asymptotic region, see for instance~\cite{dC84,nOoS11}, and in this case there is a Cauchy horizon which extends all the way to future null infinity. Although the Tolman-Bondi metrics are spherically symmetric and are known in closed explicit form, still establishing the existence of future-directed null geodesics emanating from the central singularity is a problem of considerable mathematical complexity. So far, sufficient conditions upon the initial data have been found that in turn guarantee the existence of radial null geodesics emanating from the central singularity~\cite{dC84,rN86,bNfM01,nOoS11}.

The visibility of the null singularity to local or asymptotic observers has raised questions regarding the validity of the strong and weak cosmic censorship conjectures within Einstein's general theory of relativity. It is not the intention of the present work to discuss this delicate and difficult matter, instead our intention is to provide a systematic discussion for the null geodesics (with and without angular momentum) emanating from or terminating at the central singularity.

In order to achieve this goal, here we restrict ourselves to the family of \emph{self-similar} Tolman-Bondi collapse models. For an introduction to such models, see for instance~\cite{mCaT71,kLtZ90,aOtP87,dC94,pB95,bNfM02} as well as the review in~\cite{bCaC05}, and for questions related to the stability of the Cauchy horizon and cosmic censorship based on these models see~\cite{bNfM02,bNtW02,tWbN09,eDbN11,eDbN11b}. Our motivation for restricting our attention to the self-similar case stems from the fact that the extra symmetry of the background geometry allows us to have a complete understanding on the behaviour of radial and non-radial null geodesics emanating from or terminating at the central singularity. This property is very welcome for a number of independent reasons.

Some time ago, Mena and Nolan~\cite{bNfM01} have shown that any marginally bound nakedly singular Tolman-Bondi dust collapse model admits future-directed non-radial null geodesics emanating from the central singularity. At first sight, this result might appear unexpected and counterintuitive, since by conservation of angular momentum one expects that no causal geodesics possessing non-vanishing angular momenta can pass through the center of the cloud. As long as the center is regular, this expectation is indeed fulfilled. However, when the singularity forms, the singular nature of the center transcends the restrictions imposed by the law of angular momentum conservation, as Mena and Nolan show at least for the case of marginally bound collapse. In the present work, we determine the class of all possible (radial and non-radial) null geodesics emanating from the central singularity in the self-similar case, thus complementing the results in~\cite{bNfM01,bNfM02}. For an extension of these results to the more generic family of bounded nakedly singular Tolman-Bondi models, see our recent work~\cite{nOoStZ15}.

There is another reason for which non-radial geodesics emanating from the central singularity play an important role, as we noticed in~\cite{nOoStZ15}. In that work, via a combination of analytical and numerical computations, evidence is presented to support the view that these non-radial null geodesics emanating from the singularity shape the shadow that a collapsing cloud cast upon the image perceived by an asymptotic observer of an external source illuminating the collapsing cloud. Accordingly, these geodesics may herald to an observer in the asymptotic region the formation of a Cauchy horizon.

The structure of the present paper is as follows. In the next section, we introduce the family of self-similar Tolman-Bondi spacetimes and discuss briefly their basic properties. In the following section, we discuss the important family of radially in- and outgoing null geodesics and explain the causal structure of the spacetime.  In the subsequent section, we set up a Hamiltonian formalism describing the non-radial null geodesics and give a full qualitative description of the associated Hamiltonian flow. Finally, in the last section a summary of the results is presented and possible applications are discussed.

\section{Self-similar Tolman-Bondi spacetimes }

We begin this section reviewing the self-similar Tolman-Bondi collapse. The metric describing the collapsing spacetime is
\begin{equation}
{\bf g} = -d\tau ^{2} + F^2(x) dR^2 
 + R^{2} S^{2}(x)\left( d\vartheta^2 + \sin^2\vartheta d\varphi^2 \right),
\qquad x := \frac{\tau}{R},
\label{Eq:gSS}
\end{equation}
where $\tau\in\Real$ is the proper time along the flow lines of the dust particles, $R\geq 0$ is a comoving radial coordinate labeling the dust shells, and $(\vartheta,\varphi)$ are standard angular coordinates on the two-sphere, while the functions $F$ and $S$ are defined by
\begin{equation}
F(x) := \frac {1-\frac {\lambda x}{3}}{(1-\lambda x)^{1/3}},\quad 
S(x) := (1-\lambda x)^{2/3},\qquad x < 1/\lambda, 
\label{FunFS}
\end{equation}
with $\lambda$  a positive parameter. Denoting by $m(R)$ the Misner-Sharp mass function, the parameter $\lambda$ characterizes the initial compactness ratio of each dust shell,
$$
\frac{2m(R)}{R} = \frac{4}{9}\lambda^2,
$$
which turns out to be independent of $R$ in this model. The areal radius $r(\tau,R) = RS(x)$ satisfies 
\begin{equation}
\frac {1}{2}\left( \frac{\partial r(\tau,R)}{\partial\tau} \right)^2 - \frac {m(R)}{r(\tau,R)} = 0,
\label{EqR.}
\end{equation}
implying that the collapsing shells have zero total energy. Furthermore, $\partial r/\partial R = F(x) > 0$ is positive, excluding the formation of shell-crossing singularities.

The spacetime described by Eq.~(\ref{Eq:gSS}) has a shell-focusing singularity at $x = 1/\lambda$, where $r/R$ vanishes and the density and the curvature blow up. Depending upon the range of the parameter $\lambda$, this singularity may be visible to local observers (see Lemma~\ref{Lem:Y} below). In this work, we focus on the case where the shell-focusing singularity is visible and provide a detailed analysis for the behaviour of radial and non-radial null geodesics emanating from it.

A particular property of the metric~(\ref{Eq:gSS}) is the presence of the homothetic Killing vector field
\begin{equation}
\xi = \tau\frac {\partial }{\partial \tau} + R\frac {\partial }{\partial R},
\label{Eq:HG}
\end{equation}
such that the Lie derivative of ${\bf g}$ along the integral curves of $\xi$ scales according to $\pounds_{\xi} {\bf g} = 2{\bf g}$. The existence of $\xi$ yields a conserved quantity for the equations of motion describing null geodesics:

\begin{proposition}
\label{Prop:Homothetic}
Let $\xi$ be a homothetic Killing vector field admitted by a Lorentz manifold $(M,{\bf g})$, and consider an affinely parametrized null geodesic $\gamma$ in $(M,{\bf g})$ with tangent vector field ${\bf p}$. Then, the quantity $C:={\bf g}({\bf p},\xi)$ remains constant along $\gamma$.
\end{proposition}

\proof Using local coordinates $(x^\mu)$ on $(M,{\bf g})$, we find
$$
p^\mu\nabla_\mu C = p^\mu\nabla_\mu (p^\nu \xi_\nu)
 = (p^\mu\nabla_\mu p^\nu)\xi_\nu + p^\mu p^\nu \nabla_\mu\xi_\nu.
$$
The first term on the right-hand side vanishes because ${\bf p}$ is geodesic and affinely parametrized, while the second term vanishes because $\nabla_\mu\xi_\nu + \nabla_\nu\xi_\mu = \pounds_\xi g_{\mu\nu} = 2g_{\mu\nu}$ and ${\bf p}$ is null. 
\qed

Notice that the norm of $\xi$ is
$$
{\bf g}(\xi,\xi) = -R^2(x^2 - F^2(x)),
$$
implying that $\xi$ is timelike in the region where $x^2 - F^2(x) > 0$ but spacelike where $x^2 - F^2(x) < 0$. Therefore, $C$ cannot be interpreted as a conserved ``energy" in an obvious way. Nevertheless, as we will see, the existence of the conserved quantity $C$ greatly facilitates the analysis of the geodesic flow with non-vanishing angular momentum.

\section{Radial null geodesics}

In this section, we analyze the behaviour of radial null geodesics. From Eq.~(\ref{Eq:gSS}), it follows that these geodesics are determined by
$$
d\tau = \pm F(x) dR,
$$
where the $+$ ($-$) sign refers to outgoing (ingoing) radial null geodesics. In terms of the variables $x = \tau/R$ and $s := -\log(R)$ this equation takes the form
\begin{equation}
\frac{dx}{ds} = Y_\pm(x) := x \mp F(x),\qquad x < 1/\lambda.
\label{Eq:Rad}
\end{equation}
The qualitative behaviour of the solutions of these equations can be understood from the properties of the functions $Y_\pm$, which are summarized in the following lemma.

\begin{figure}[h!]
\begin{center}
\includegraphics[width=6.0cm]{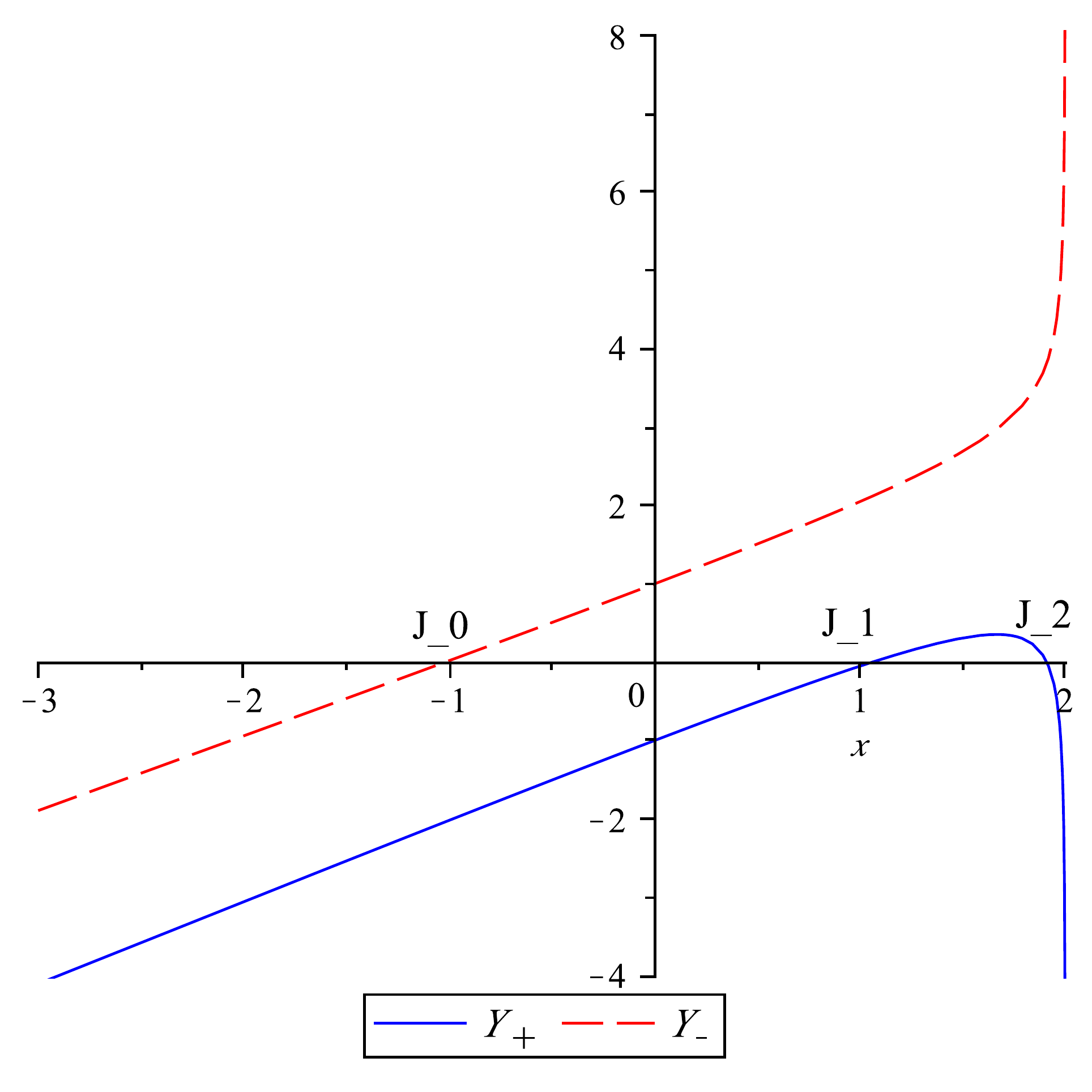}
\end{center}
\caption{\label{Fig:Ypm} A plot of the two functions $Y_+$ (blue-continuous) and $Y_-$ (red-dashed) for the parameter value $\lambda = 0.5$. In this case, $J_0 \simeq -1.0198$, $J_1Ê\simeq 1.0588$ and $J_2 \simeq 1.9088$.}
\end{figure}

\begin{lemma}
\label{Lem:Y}
The functions $Y_{\pm}: (-\infty,1/\lambda) \to \Real$ defined in Eq.~(\ref{Eq:Rad}) satisfy the following properties (cf. Fig.~\ref{Fig:Ypm}):
\begin{enumerate}
\item[(a)] $\lim\limits_{x\to -\infty} Y_{\pm}(x) = -\infty$, $\lim\limits_{x\to 1/\lambda} Y_{\pm}(x) = \mp \infty$.
\item[(b)] $Y_-$ has a single root $J_0$ satisfying $J_0 < 0$.
\item[(c)] Regarding the function $Y_+$, let
$$
\lambda^* := \frac{3}{2}(\sqrt{3}-1)(3\sqrt{3} - 5)^{1/3} \simeq 0.638.
$$
For $\lambda > \lambda^*$, $Y_+$ is strictly negative, for $\lambda = \lambda^*$, $Y_+$ has a single root at $J_1 = (\sqrt{3} - 1)(3\sqrt{3} - 5)^{-1/3}$, while for  $0 < \lambda < \lambda^*$, $Y_+$ has precisely two roots $J_1$, $J_2$ satisfying $0 < J_1 < J_2 < 1/\lambda$. 
\end{enumerate}
\end{lemma}

\proof (a) follows by inspection. For (b) and (c) the following expressions are helpful:
$$
F'(x) = \frac{2}{9}\lambda^2 x(1- \lambda x)^{-4/3},\qquad
F''(x) = \frac{2}{9}\lambda^2\left(1 +  \frac{\lambda}{3} x \right)(1 - \lambda x)^{-7/3}.
$$
In order to prove (b), we first note that the existence of the root follows from the intermediate value theorem. There are no positive roots because $Y_-(0) = 1$ and $Y_-$ is increasing in the interval $(0,1/\lambda)$. If $J_0 < 0$ is a negative zero of $Y_-$, then it is not difficult to show that $Y_-'(J_0) > 0$, implying the uniqueness of $J_0$.

Finally, in order to prove (c), we note that $Y_+(0) = -1$, $Y_+'(0) = 1$, and that $Y_+$ is increasing for $x\leq 0$ and concave for $x\in [0,1/\lambda)$. Therefore, $Y_+$ has a global maximum at some point $J_{max}\in (0,1/\lambda)$ and depending on whether the value of $Y_+$ at this maximum is negative or not, the function $Y_+$ has no roots, or it has precisely two roots $J_2\geq J_1$ in the interval $(0,1/\lambda)$. If $J_1$ is a root of $Y_+$ then, $z_1 := \lambda J_1\in (0,1)$ satisfies $z_1 = \lambda(1-z_1/3)(1-z_1)^{-1/3}$,
and
$$
\lambda = P(z_1) := \frac{z_1(1-z_1)^{1/3}}{1 - \frac{z_1}{3}}.
$$
The positive function $P: (0,1)\to \Real$ converges to zero for $z\to 0$ or $z\to 1$ and it has a maximum at $z_1 = z_1^* = 3(2-\sqrt{3})$, where
$$
\lambda^* = P(z_1^*) = \frac{3}{2}(\sqrt{3}-1)(3\sqrt{3} - 5)^{1/3}.
$$
Therefore, the function $Y_+$ has no roots when $\lambda > \lambda^*$, it has a degenerate root at $J_1 = J_2 = z_1^*/\lambda^* = (\sqrt{3} - 1)(3\sqrt{3} - 5)^{-1/3}$ when $\lambda = \lambda^*$, and it has two roots when $0 < \lambda < \lambda^*$.
\qed

\begin{figure}[h!]
\begin{center}
\includegraphics[height=6.5cm]{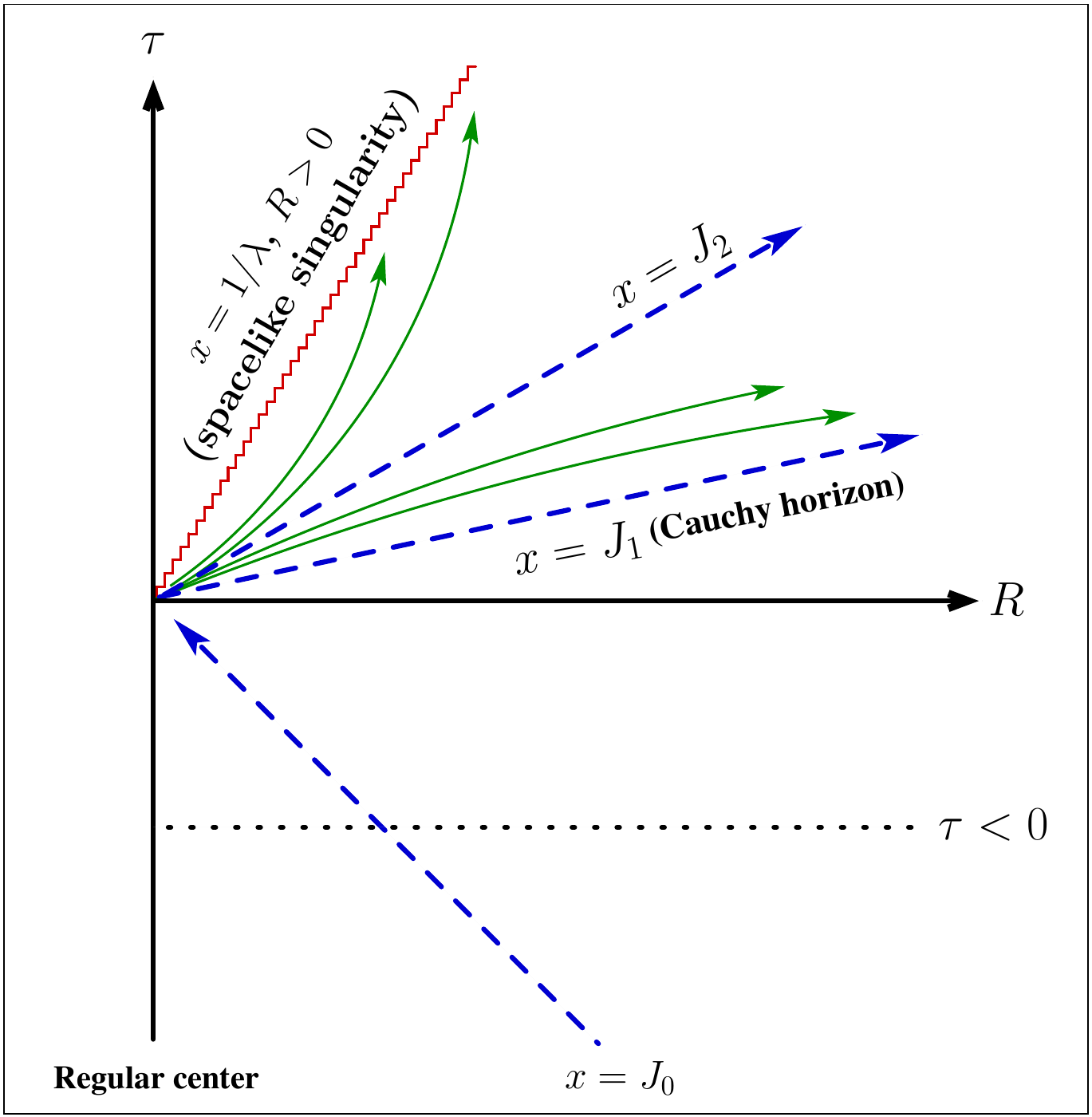}
\hspace{0.5cm}
\includegraphics[height=6.5cm]{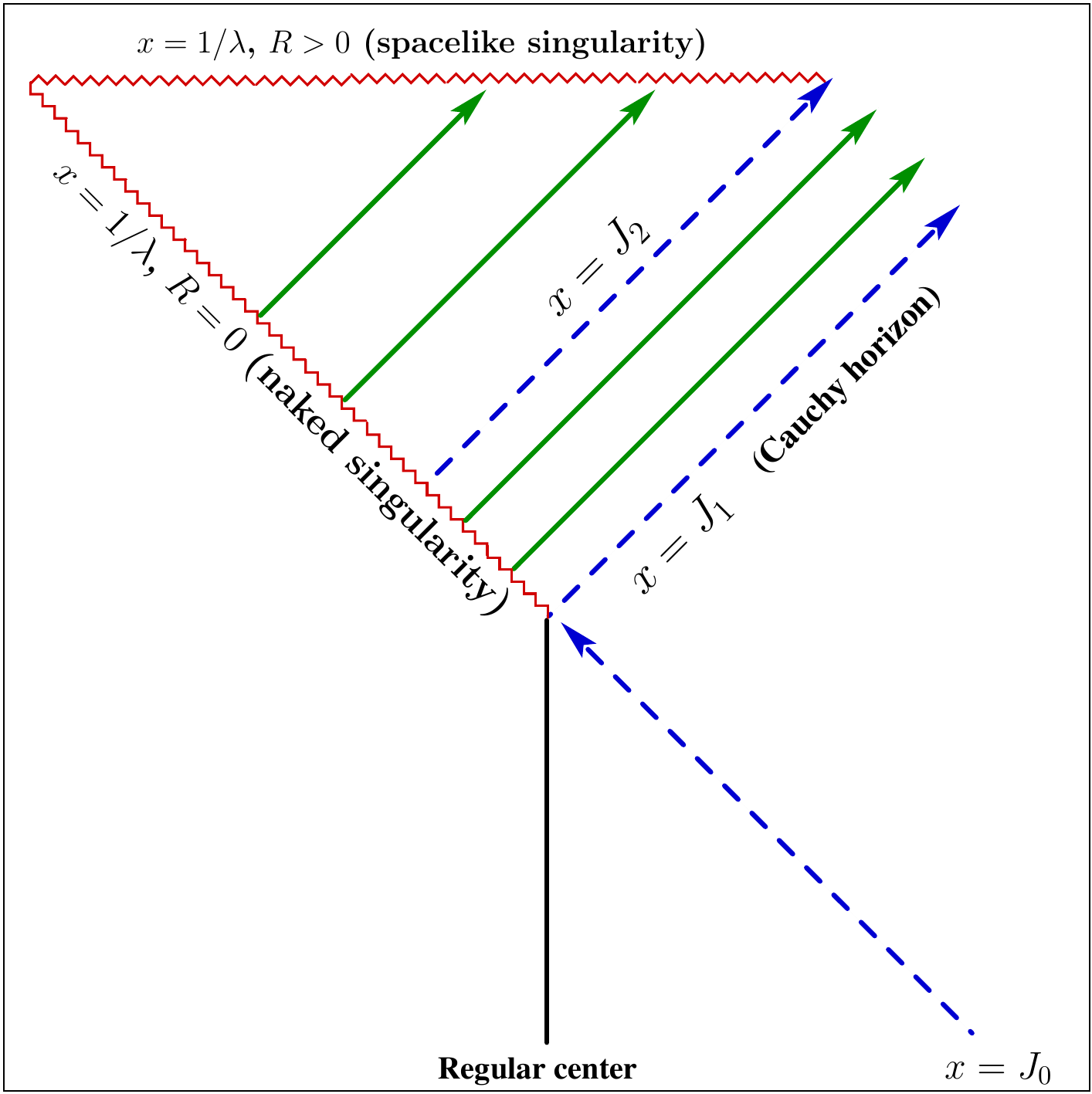}
\end{center}
\caption{\label{Fig:Conformal} Spacetime diagrams illustrating the causal structure of self-similar dust collapse. Left panel: The location of the singularity and the particular radial null rays $x = J_0,J_1,J_2$ in the $\tau-R$ plane. The green solid lines with the arrows represent future-directed radial null geodesics emanating from the central singularity. Right panel: The location of the same objects in the conformal diagram.}
\end{figure}

For the following, we assume that the parameter $\lambda$ in Eq.~(\ref{FunFS})  has been chosen so that it lies below the critical value $\lambda^*$ i.e. $0 < \lambda < \lambda^*$,
and thus  the function $Y_+$ has precisely two zeros denoted by $ J_1 < J_2$. Consequently, in the outgoing case, the system~(\ref{Eq:Rad}) has two critical points at $x = J_1$ and $x = J_2$. It follows from the behaviour of $Y_+$ that $x = J_1$ is unstable, while $x = J_2$ is an attractor: any solution with initial data $x(0)\in (J_1,1/\lambda)$ converges to $J_2$ as $s\to \infty$ ($R\to 0$), while any solution with $x(0)\in (-\infty,J_1)$ converges to $-\infty$ when $s\to \infty$ ($R\to 0$). The solution is implicitly determined by the equation
\begin{equation}
s - s_0 = \int\limits_{x(0)}^{x(s)} \frac{dx}{x - F(x)}.
\label{Eq:RadGeoOut}
\end{equation}
It follows that any outgoing radial null geodesic that passes through a point $(\tau,R)$ such that $J_1\leq \tau/R < 1/\lambda$, $R > 0$, emanates from the central singularity $(\tau,R) = (0,0)$, see Fig.~\ref{Fig:Conformal}. On the other hand, the outgoing radial null geodesics passing through a point $(\tau,R)$ with $\tau/R < J_1$ and $R > 0$ emanate from a point $(\tau_0 < 0,0)$ on the regular center.

Regarding the case of the ingoing radial null geodesics, it follows from Eq.~(\ref{Eq:Rad}) and the behaviour of $Y_-$, that $x = J_0$ is an unstable critical point. This implies that any ingoing radial null geodesic passing through a point $(\tau,R)$ with $\tau/R < J_0$ and $R > 0$ terminates at the regular center, while any ingoing radial null geodesic passing through a point $(\tau,R)$ with $\tau/R > J_0$ and $R > 0$ terminates at the shell-focussing singularity with finite value of $R$, as can be seen from the convergence of the integral
\begin{equation}
s - s_0 = \int\limits_{x(0)}^{x(s)} \frac{dx}{x + F(x)}.
\label{Eq:RadGeoIn}
\end{equation}
when $x(0) > J_0$ and $x(s)\to 1/\lambda$.

Therefore, there is a unique ingoing radial null geodesic terminating at the central singularity $(\tau,R) = (0,0)$, while there are infinitely many outgoing radial null geodesics emanating from it. The earliest one, $x = J_1$, describes the Cauchy horizon, see Fig.~\ref{Fig:Conformal}.

\section{Hamiltonian description of the null geodesic flow}

In the previous section, we reviewed the behaviour of radial in- and outgoing null geodesics. In this section, we shall appeal to the powerful Hamiltonian formalism to treat the more general case of non-radial null geodesics in the background metric described in Eq.~(\ref{Eq:gSS}). Via this formalism, based on the rotational symmetry and the presence of the homothetic Killing vector field, we show that the description of the complete set of null geodesics is reduced to an effective one-dimensional Hamiltonian system.

We recall that geodesic motion is described by the Hamiltonian
$$
H(x,p) = \frac{1}{2} g^{\mu\nu}(x) p_\mu p_\nu,
$$
and the null geodesics are those trajectories along which $H = 0$. Because of spherical symmetry, it is sufficient to restrict ourselves to the equatorial plane $\vartheta = \pi/2$ and moreover it is convenient to express $H(x,p)$ in terms of the local coordinates $(x,s,\vartheta,\varphi)$ where $x = \tau/R$ and $s = -\log(R)$. Relative to these coordinates, the homothetic Killing vector field assumes the simple form
$$
\xi = -\frac{\partial}{\partial s},
$$
and the inverse metric on the equatorial plane reads
$$
{\bf g}^{-1} = e^{2s}\left[ -\frac{\partial}{\partial x}\otimes\frac{\partial}{\partial x}
 + \frac{1}{F^2(x)}\left( \frac{\partial}{\partial s} + x \frac{\partial}{\partial x} \right)
 \otimes \left( \frac{\partial}{\partial s} + x \frac{\partial}{\partial x} \right)
  + \frac{1}{S^2(x)}\frac{\partial}{\partial \varphi}\otimes\frac{\partial}{\partial \varphi}
  \right].
$$
Since conformal transformations do not affect the null geodesics as trajectories, hereafter, we discard the conformal factor $e^{2s}$. The resulting Hamiltonian takes the form
$$
\tilde{H}(s,x,\varphi,p_s,p_s,p_\varphi)
 = \frac{1}{2}\left[ -p_x^2 + \frac{(p_s + x p_x)^2}{F^2(x)} + \frac{p_\varphi^2}{S^2(x)} \right].
$$
The variables $s$ and $\varphi$ are cyclic, and thus the following two quantities
\begin{equation}
C := -p_s,\qquad
\ell := p_\varphi,
\end{equation}
are preserved along the flow. The quantity $C$ is the constant of motion associated with the homothetic Killing vector field $\xi$, see Proposition~\ref{Prop:Homothetic}, while $\ell$ is the constant of motion associated with the rotational Killing vector field $\partial/\partial\varphi$. As a consequence of these remarks, the Hamiltonian can be reduced to the effective one-dimensional system
\begin{equation}
\tilde{H}_{C,\ell}(x,p_x) 
 =  \frac{1}{2}\left[ -p_x^2 + \frac{(x p_x - C)^2}{F^2(x)} + \frac{\ell^2}{S^2(x)} \right].
\end{equation}
This effective Hamiltonian describes the motion in the $x-p_x$ plane:
\begin{eqnarray}
\dot{x} &=& \frac{\partial \tilde{H}_{C,\ell}}{\partial p_x}
 = -p_x + \frac{x}{F^2(x)}(x p_x - C),
\label{Eq:xdot}\\
\dot{p_x} &=& -\frac{\partial \tilde{H}_{C,\ell}}{\partial x} 
 = -\frac{p_x}{F^2(x)}(x p_x - C) + \frac{(xp_x - C)^2}{F^3(x)} F'(x)
 + \frac{\ell^2}{S^3(x)} S'(x),
\end{eqnarray}
and we are only interested in those trajectories for which $\tilde{H}_{C,\ell} = 0$. Eliminating $p_x$ in Eq.~(\ref{Eq:xdot}) by using $\tilde{H}_{C,\ell} = 0$, we obtain
\begin{equation}
\dot{x}^2 + V_{C,\ell}(x) = 0,\qquad
V_{C,\ell}(x) := \frac{1}{F^2(x) S^2(x)}\left[ \ell^2(x^2 - F^2(x)) - C^2 S^2(x) \right],
\label{Eq:1D}
\end{equation}
which is the equation for a point particle with zero energy moving in one dimension.

Once the motion in the $x-p_x$ plane has been determined, the additional degrees of freedom describing the flow can be obtained by integrating the equations
\begin{eqnarray}
\dot{s} &=& \frac{\partial \tilde{H}}{\partial p_s} = \frac{x p_x - C}{F^2(x)},
\label{Eq:sdot}\\
\dot{\varphi} &=& \frac{\partial \tilde{H}}{\partial p_\varphi} = \frac{\ell}{S^2(x)}.
\label{Eq:phidot}
\end{eqnarray}

\subsection{The radial case ($\ell=0$)}

As a first application of the formalism developed in this section, here we briefly reconsider the case of radial null geodesics. The equations for these geodesics are obtained by setting  $\ell=0$ in Eq.~(\ref{Eq:1D}) which yields $\dot{x} = \pm C/F(x)$. On the other hand, the constraint $\tilde{H}_{C,\ell} = 0$ yields
$$
\left[ (x - F(x))p_x - C \right]\left[(x + F(x))p_x - C \right] = 0.
$$
Assume first that $F^2(x)\neq x^2$. Then, we obtain from this the solutions
$$
p_x = \frac{C}{x \mp F(x)}.
$$
Introducing this relation into Eqs.~(\ref{Eq:xdot},\ref{Eq:sdot}) we obtain
\begin{equation}
\dot{x} = \pm \frac{C}{F(x)},\qquad
\dot{s} = \pm \frac{C}{F(x)}\frac{1}{x\mp F(x)}.
\label{Eq:xsdotRad}
\end{equation}
If $C=0$, then $\dot{x}=\dot{s}=0$ and $p_x = p_s = p_\varphi = 0$, so we obtain a trivial geodesic. If $C\neq 0$, we can divide $\dot{x}$ by $\dot{s}$ and we recover Eq.~(\ref{Eq:Rad}).

When $F^2(x) = x^2$, we obtain from $\tilde{H}_{C,\ell} = 0$ either $C = 0$ or $p_x = C/(2x)$. In the first case, $\dot{x} = 0$ and $\dot{s} = p_x/x$, and thus we recover the critical radial null geodesics $x = J_a$, $a=0,1,2$ discussed in the previous section. In the second case, we obtain $\dot{x} = -C/x$ and $\dot{s} = -C/(2x^2)$, which is consistent with the limit of Eq.~(\ref{Eq:xsdotRad}) when $F(x)\to \mp x$.

\subsection{The non-radial case with $C = 0$}

When $C = 0$ and $\ell\neq 0$, it follows from Eq.~(\ref{Eq:1D}) that the motion is restricted to the region where $x^2 - F^2(x)\leq 0$, implying that $\xi$ is spacelike or null, and in this case, the Hamiltonian constraint $\tilde{H}_{C,\ell} = 0$ simplifies to
$$
(x^2 - F^2(x)) p_x^2 + \ell^2\frac{F^2(x)}{S^2(x)} = 0.
$$
Since $\ell^2 F^2/S^2$ is positive, neither $x^2 - F^2(x)$ nor $p_x$ can vanish, and it follows that $\xi$ is spacelike and that the motion is confined to the regions $J_0 < x < J_1$ or $J_2 < x < 1/\lambda$. Dividing Eq.~(\ref{Eq:sdot}) by Eq.~(\ref{Eq:xdot}) we obtain
$$
\frac{ds}{dx} = -\frac{x}{F^2(x) - x^2}.
$$
Integration leads to
\begin{equation}
s = s_0 - \int\limits_{x_0}^{x(s)} \frac{x dx}{F^2(x) - x^2},
\end{equation}
which determines the trajectory of the light ray in the $x-s$ plane, for given initial data $(x_0,s_0)$ with $s_0\in\Real$ and $x_0$ lying either in the interval $(J_0,J_1)$ or in the interval $(J_2,1/\lambda)$. In the first case, $s\to -\infty$ $(R\to\infty)$ as $x(s)\to J_0$ or $x(s)\to J_1$, and thus we obtain a null geodesic which asymptotes to the Cauchy horizon $x = J_1$ in the future and to $x = J_0$ in the past. In the second case, $s\to \infty$ ($R\to 0$) as $x(s)\to J_2$, while $s$ and $R$ converge to a finite value when $x(s)\to 1/\lambda$, so in this case we have a null geodesic emanating from the central singularity which terminates at the spacelike portion of the singularity.

\subsection{The generic case ($\ell\neq 0$, $C\neq 0$)}

Finally, we treat the generic case of non-radial null geodesics with constants of motion $C$ and $\ell$ different from zero. In this case, the motion in the $x$-direction is restricted to the set of points for which (see Eq.~(\ref{Eq:1D}))
\begin{equation}
W(x) := -\frac{{\bf g}(\xi,\xi)}{r^2} = \frac{x^2 - F^2(x)}{S^2(x)} \leq \frac{1}{\beta^2},
\label{Eq:WDef}
\end{equation}
where we have introduced the ``impact parameter" $\beta := \ell/C$. The structure of this set is discussed in the following lemma.

\begin{lemma}
\label{Lem:W}
Consider the effective potential $W : (-\infty,1/\lambda)\to \Real$ defined by Eq.~(\ref{Eq:WDef}).  For each $\beta\neq 0$ let $I_\beta$ be the set consisting of those points $x$ for which $W(x) < 1/\beta^2$. Then, there exists $\beta_c > 0$ depending on $\lambda$ such that\begin{equation}
I_\beta = \left\{ \begin{array}{ll}
(x_0,1/\lambda)\hbox{ for some $x_0 < J_0$}, &  \beta^2 < \beta_c^2,\\
(x_0,1/\lambda)\setminus\{ x_1 \} \hbox{ for some $x_0 < J_0$ and $J_1 < x_1 < J_2$},
& \beta^2 = \beta_c^2,\\
(x_0,x_1) \cup (x_2,1/\lambda) \hbox{ for some $x_0 < J_0$ and $J_1 < x_1 < x_2 < J_2$},
& \beta^2 > \beta_c^2
\end{array}Ê\right.
\end{equation}
The allowed region for the motion in the $x$-direction is the closure, $\overline{I_\beta}$ in $(-\infty,1/\lambda)$, of the set $I_\beta$.
\end{lemma}

\proof The function $W: (-\infty,1/\lambda)\to \Real$ satisfies $\lim_{x\to -\infty} W(x) = +\infty$ and $\lim_{x\to 1/\lambda} W(x) = -\infty$, and as a consequence of Lemma~\ref{Lem:Y} it is positive on the intervals $(-\infty,J_0)$ and $(J_1,J_2)$ and negative on the intervals $(J_0,J_1)$ and $(J_2,1/\lambda)$, with simple roots at $J_0,J_1,J_2$. Furthermore, a straightforward calculation reveals that
\begin{equation}
W'(x) = 2\left( 1 - \frac{\lambda x}{3}Ê\right)(1 - \lambda x)^{-7/3}
\left[ x - \frac{2\lambda}{3}(1-\lambda x)^{-2/3} \right],
\label{Eq:dW}
\end{equation}
implying that $W$ is decreasing for $x < 0$. Hence, for any $\beta^2 > 0$ there exists a unique $x_0 < J_0$ such that $1/\beta^2 = W(x_0)$. Therefore, $I_\beta \subset (x_0,1/\lambda)$. If $\beta^2$ is small enough, it is clear that the inequality~(\ref{Eq:WDef}) is satisfied for all $x\in (x_0,1/\lambda)$, because $W(x) \to -\infty$ as $x\to 1/\lambda$, so in this case $I_\beta = (x_0,1/\lambda)$. On the other hand, for large $\beta^2$, the inequality Eq.~(\ref{Eq:WDef}) cannot hold for all $x\in (J_1,J_2)$ since $W$ is positive on this interval.

We now claim that the function $W'$ has a unique root $x_c$ on the interval $(J_1,J_2)$, corresponding to a maximum of $W$ on $(J_1,J_2)$. The statement of the lemma then follows with $\beta_c := W(x_c)^{-1/2}$. In order to prove the claim, we consider the function $U: (J_1,J_2)\to \Real$, $x\mapsto x - (2\lambda/3)(1-\lambda x)^{-2/3}$ which determines the sign of $W'(x)$, see Eq.~(\ref{Eq:dW}). From the behaviour of the function $x^2 - F^2(x)$ that follows from Lemma~\ref{Lem:Y}, we know that $W'(J_1) > 0$ and $W'(J_2) < 0$, implying that $U$ is positive near $x = J_1$ and negative near $x = J_2$. By the intermediate value theorem, there exists a point $x_1 \in (J_1,J_2)$ such that $U(x) > 0$ for all $x\in (J_1,x_1]$ and $U'(x_1) < 0$. Since $U$ is concave, it follows that $U'(x) < 0$ for all $x\in (x_1,J_2)$ which implies that $U$ has a unique zero on $(J_1,J_2)$. This concludes the proof of the lemma.
\qed

Restricting ourselves to the allowed region $\overline{I_\beta}$, we use $\tilde{H}_{C,\ell} = 0$ to obtain
\begin{equation}
p_x = \left\{ \begin{array}{ll}
 \frac{C}{x^2 - F^2(x)}\left[ x \pm F(x) Q_\beta(x) \right], & F^2(x) \neq x^2,\\
 \frac{C}{2x}\left[ 1 + \beta^2\frac{x^2}{S^2(x)} \right], & F^2(x) = x^2
\end{array} \right.
\end{equation}
with
$$
Q_\beta(x) := \sqrt{1 - \beta^2 W(x)} = \sqrt{1 - \frac{\beta^2}{S^2(x)}(x^2 - F^2(x))}.
$$
In the first case, when $F^2(x) \neq x^2$, we obtain from Eqs.~(\ref{Eq:xdot},\ref{Eq:sdot}),
\begin{equation}
\dot{x} = \pm \frac{C}{F(x)} Q_\beta(x),\qquad
\dot{s} = \pm \frac{C}{F(x)} \frac{x Q_\beta(x) \pm F(x)}{x^2 - F^2(x)}.
\label{Eq:xsdotNonRad1}
\end{equation}
In the second case,
\begin{equation}
\dot{x} = -\frac{C}{x},\qquad
\dot{s} = -\frac{C}{2x^2}\left( 1 - \beta^2\frac{x^2}{S^2(x)} \right),
\label{Eq:xsdotNonRad2}
\end{equation}
which is seen to be the limit of Eq.~(\ref{Eq:xsdotNonRad1}) when $F(x)\to \mp x$. We also see that Eq.~(\ref{Eq:xsdotNonRad1}) reduces to the corresponding equation~(\ref{Eq:xsdotRad}) in the radial case when $\beta \to 0$.

\begin{figure}[h!]
\begin{center}
\includegraphics[width=7.0cm]{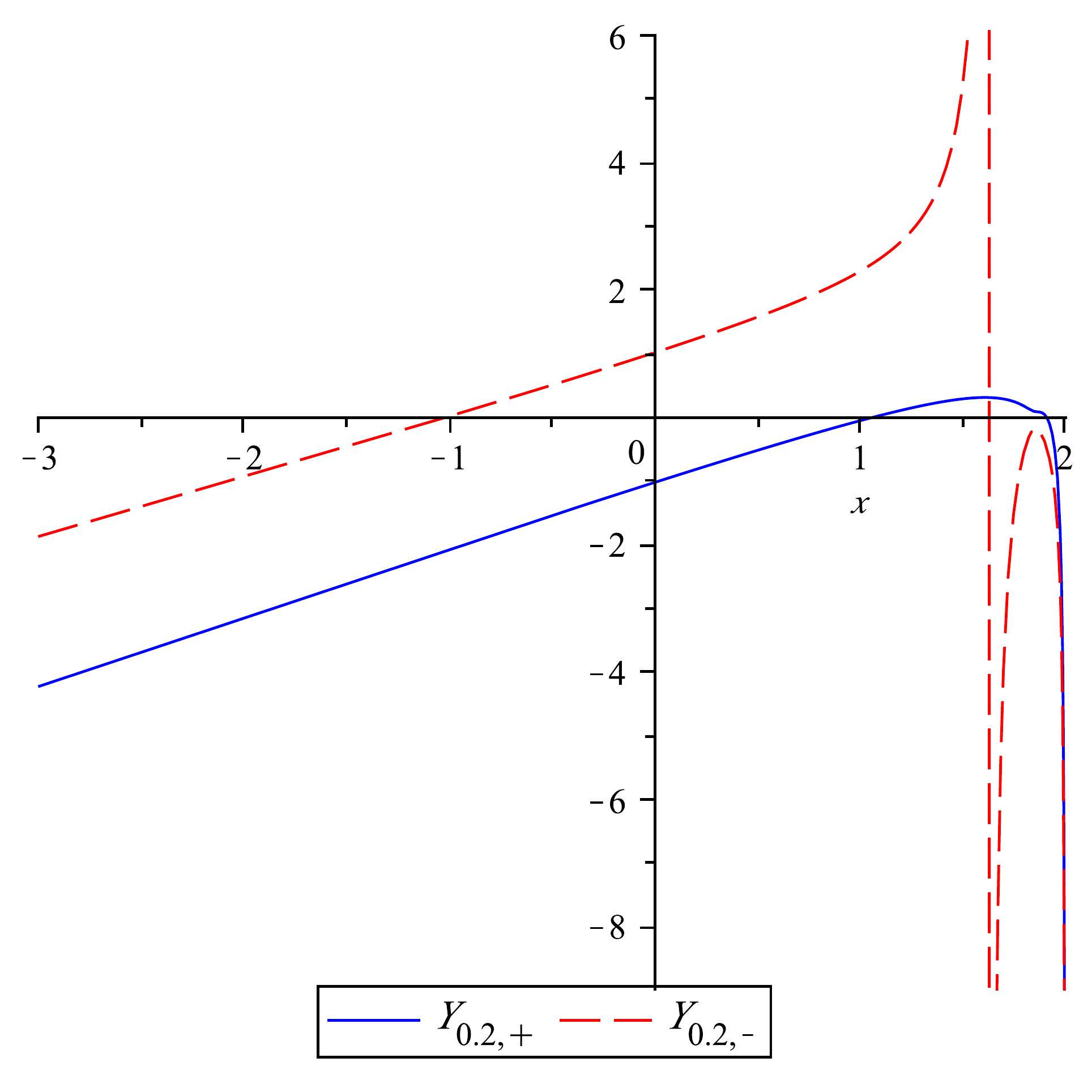}
\includegraphics[width=7.0cm]{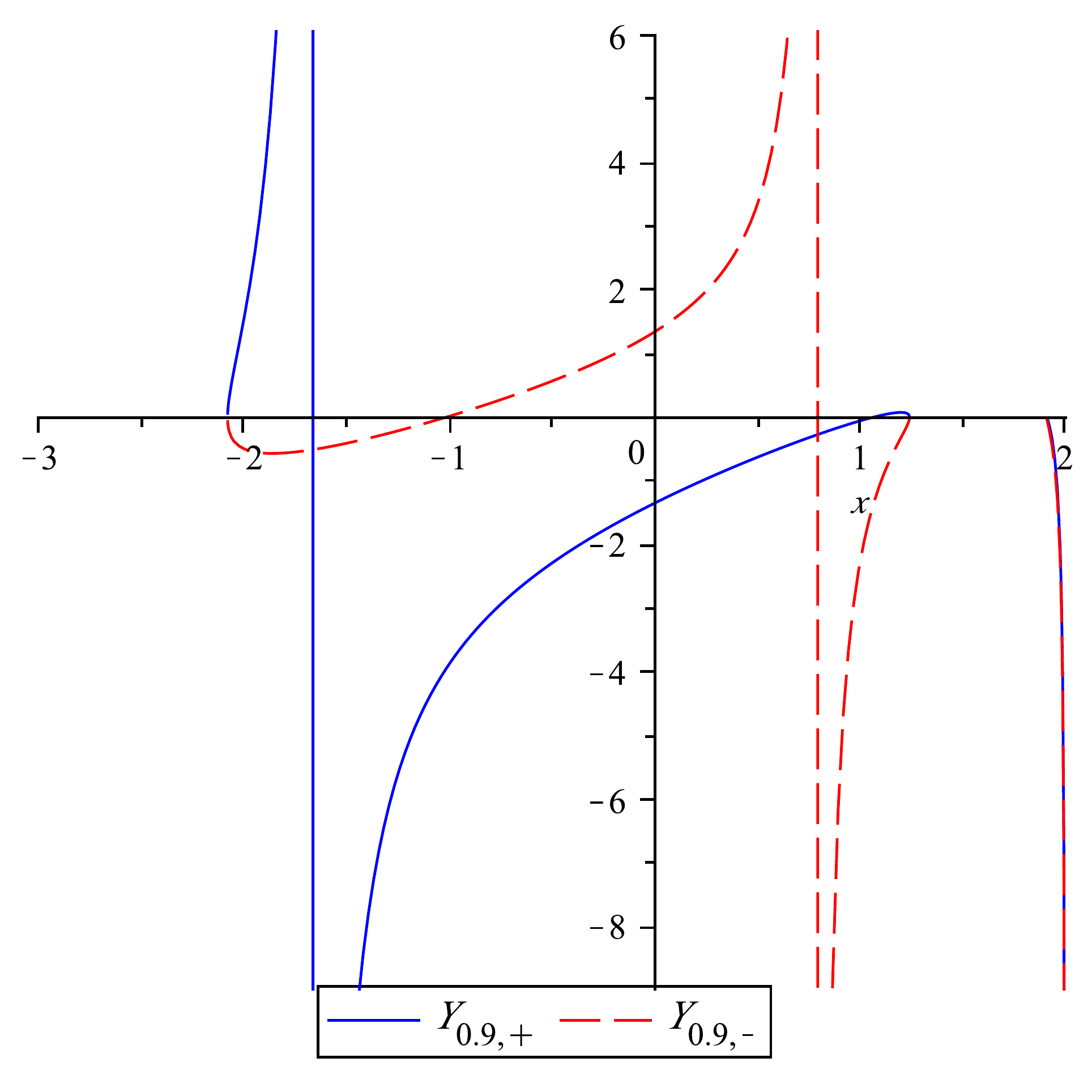}
\end{center}
\caption{\label{Fig:Ybetapm} A plot of the two functions $Y_{\beta,+}$ and $Y_{\beta,-}$ for the parameter value $\lambda = 0.5$. Left panel: $\beta = 0.2$, which lies below the critical value $\beta_c$. The two graphs connect at the point $x_0$ which lies around $x = -35.2$. Right panel: $\beta = 0.9$, which lies above the critical value $\beta_c$. In this case, it is clearly visible how the graphs of the two functions $Y_{0.9,\pm}$ connect to each other at the turning points $x = x_0$, $x = x_1$ and $x = x_2$.}
\end{figure}

For the analysis below, the following identity is worth noticing:
$$
(x Q_\beta(x))^2 - F^2(x) = (x^2 - F^2(x))\left( 1 - \beta^2\frac{x^2}{S^2(x)} \right).
$$
Using this identity and Eqs.~(\ref{Eq:xsdotNonRad1},\ref{Eq:xsdotNonRad2}) we obtain
\begin{equation}
\frac{dx}{ds} = Y_{\beta,\pm}(x) 
 := \frac{Q_\beta(x)}{1 - \beta^2\frac{x^2}{S^2(x)}}(x Q_\beta(x) \mp F(x))
 = Q_\beta(x) \frac{x^2 - F^2(x)}{xQ_\beta(x) \pm F(x)}.
\label{Eq:Ybetapm}
\end{equation}
The functions $Y_{\beta,\pm}$ determine the trajectories of non-radial null geodesics in the $\tau-R$ plane. Note that these functions reduce to the functions $Y_\pm$ describing radial null geodesics when $\beta\to 0$, see Eq.~(\ref{Eq:Rad}). The qualitative properties of the functions $Y_{\beta,\pm}$ are summarized in the next lemma and illustrated in Fig.~\ref{Fig:Ybetapm}.

\begin{lemma}
\label{Lem:Ybetapm}
Denote by $D_1 < 0 < D_2$ the two roots of the equation $\beta^2 x^2 = S^2(x)$. Then, $Y_{\beta,+}$ in Eq.~(\ref{Eq:Ybetapm}) yields a well-defined smooth function $Y_{\beta,+}: I_\beta\setminus \{ D_1 \} \to \Real$. It has a first-order pole at $x = D_1$, is positive on $(x_0,D_1) \cup (J_1,J_2)$ and negative on $(D_1,J_1)\cup (J_2,1/\lambda)$.

Similarly, $Y_{\beta,-}$ in Eq.~(\ref{Eq:Ybetapm}) yields a well-defined smooth function $Y_{\beta,-}: I_\beta\setminus \{ D_2 \} \to \Real$ which has a first-order pole at $x = D_2$, is negative on $(x_0,J_0) \cup (D_2,1/\lambda)$ and positive on $(J_0,D_2)$.
\end{lemma}

{\bf Remark}: Note that $D_1$ and $D_2$ lie inside the region $I_\beta$, since
$$
W(D_a) = \frac{1}{\beta^2} - \frac{F^2(D_a)}{S^2(D_a)} < \frac{1}{\beta^2},\quad a=1,2.
$$

\proof The proof of Lemma~\ref{Lem:Ybetapm} follows directly from the two representations of $Y_{\beta,\pm}$ in Eq.~(\ref{Eq:Ybetapm}) and the known behaviour of the function $x^2 - F^2(x)$.
\qed

From the qualitative behaviour of the functions $Y_{\beta,\pm}$ described in Lemma~\ref{Lem:Ybetapm} one finds the following properties of the null geodesics: consider first a solution of $dx/ds = Y_{\beta,+}(x)$ with initial data at $(x,s) = (x_0,s_0)$, where $x_0$ is the turning point defined in Lemma~\ref{Lem:W}. This solution extends through the pole at $x = D_1$ to $x = J_1$. Since
$$
s - s_0 = \int\limits_{x_0}^{x(s)} \frac{dx}{Y_{\beta,+}(x)},
$$
the pole of $Y_{\beta,+}$ simply describes a turning point of $s$ (or $R$). Thus, this solution describes a null geodesic emanating from the point $(x_0,s_0)$ which asymptotes to the Cauchy horizon in the future. To describe the past of this geodesic, one has to resort to the function $Y_{\beta,-}$ since $x_0$ is a turning point for the motion in the $x$-direction. From the properties of $Y_{\beta,-}$, one sees that the null geodesic asymptotes to the light ray $x = J_0$ in the past.

\begin{figure}[h!]
\begin{center}
\includegraphics[width=7.0cm]{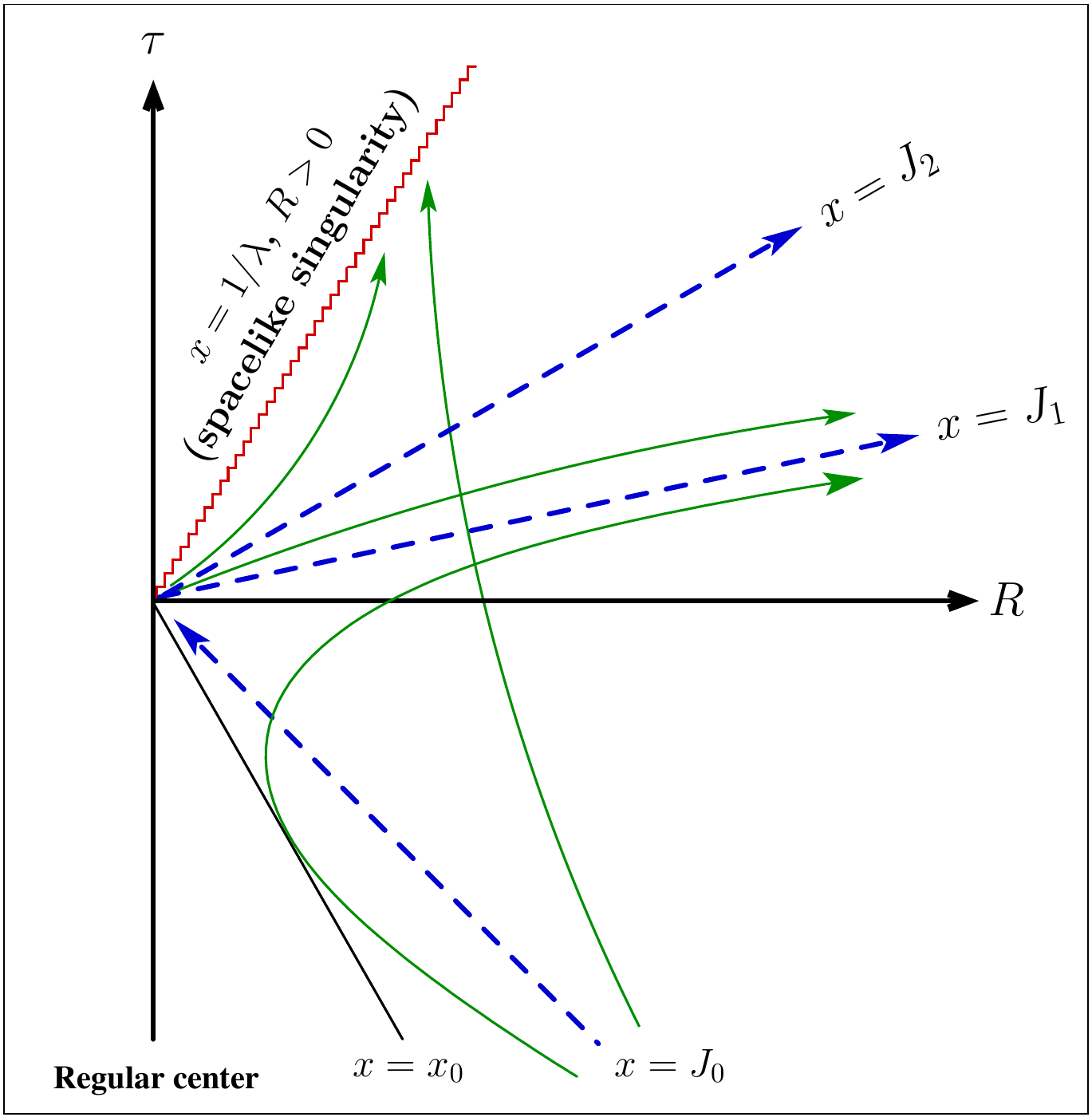}
\hspace{1cm}
\includegraphics[width=7.0cm]{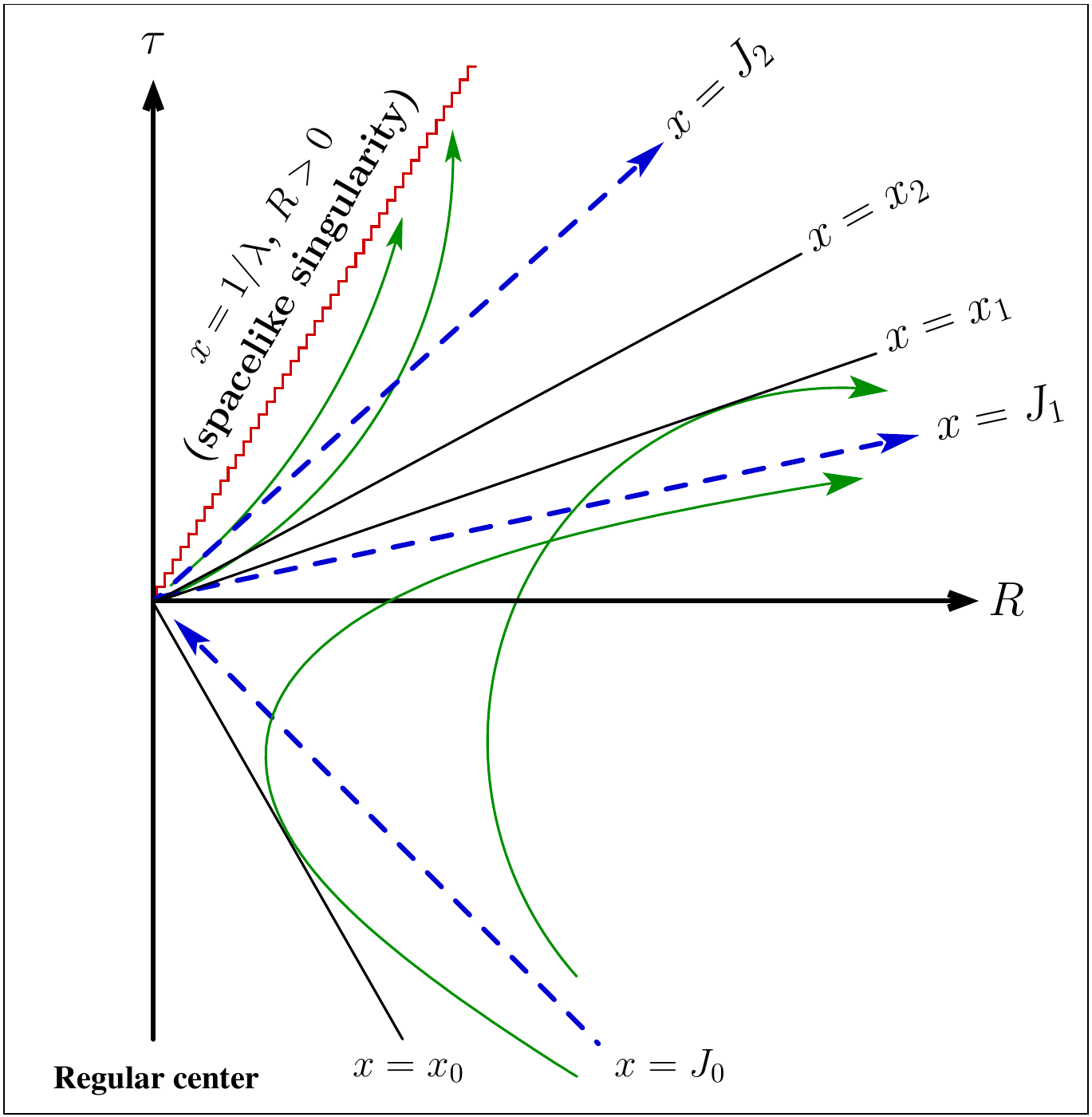}
\end{center}
\caption{\label{Fig:NonradialNullGeodesics} $\tau-R$ plane showing future-directed non-radial null light rays (green lines with arrows) propagating in the spacetime describing self-similar dust collapse. Left panel: $\beta^2 < \beta_c^2$. Right panel: $\beta^2 > \beta_c^2$.}
\end{figure}

Next, consider a light ray described by $Y_{\beta,+}$ passing through a point $(x,s)$ with $J_1 < x < 1/\lambda$. For $\beta^2 < \beta_c^2$ the picture looks qualitatively the same as in the radial case. In particular, all such null geodesics emanate from the central singularity. Regarding the null geodesics described by $Y_{\beta,-}$ passing through a point $(x,s)$ with $J_0 < x < 1/\lambda$, they asymptote to $x = J_0$ in the past, and terminate at the spacelike portion of the singularity, see Fig.~\ref{Fig:NonradialNullGeodesics}.

Therefore, when $\beta^2 < \beta_c^2$, we have the following properties: all the null geodesics passing through a point $(x,s)$ with $x_0\leq x < J_0$ bounce off at $x = x_0$ and asymptote to the Cauchy horizon in the future, while null geodesics passing through a point $(x,s)$ with $J_0 < x < J_1$ terminate at the spacelike portion of the singularity. Finally, null geodesics passing through a point $(x,s)$ with $J_1 < x < 1/\lambda$ emanated from the central singularity $(\tau,R) = (0,0)$.

For $\beta^2 > \beta_c^2$, as in the previous case, there are null geodesics which asymptote to $x = J_0$ in the past direction and to $x = J_1$ in the future direction, bouncing off at $x = x_0$. However, the qualitative features of the remaining null geodesics, namely those that penetrate or lie to the future of the Cauchy horizon are different than in the previous case: the effect of having a high angular momentum such that $\beta^2 > \beta_c^2$ is the appearance of the ``forbidden" region $x_1 < x < x_2$ (see Lemma~\ref{Lem:W}), which prevents the null geodesics originating from the far past $\tau\to -\infty$ to reach the singularity. As $\beta^2$ increases, the gap between $x = x_1$ and $x = x_2$ increases, until $x_1\to J_1$ and $x_2 \to J_2$, see Eq.~(\ref{Eq:WDef}).

Finally, we comment on the behaviour of the azimuthal angle $\varphi$ for the null geodesics emanating from the central singularity. In the cases observed so far, where $0 <\beta^2\neq \beta_c^2$, these geodesics have $x\to J_2$ as they approach the singularity. It follows from Eqs.~(\ref{Eq:phidot},\ref{Eq:xsdotNonRad1}) that
$$
\frac{d\varphi}{dx} = \pm \frac{\beta}{S^2(x)}\frac{F(x)}{Q_\beta(x)},
$$
implying that $\varphi$ has a finite limit as $x\to J_2$. However, notice that in the particular case where $\beta = \pm \beta_c$ there exist null geodesics confined to the timelike surface $x = x_c = const.$, the value $x_c$ corresponding the the local maximum of the effective potential $W$. In this particular case, it follows from Eqs.~(\ref{Eq:phidot},\ref{Eq:xsdotNonRad1}) that
$$
\frac{d\varphi}{ds} = \pm \frac{1}{\beta_c},
$$
implying that $|\varphi|\to \infty$ as the central singularity is approached. This situation is somehow analogue to the existence of unstable circular null geodesics in the Schwarzschild spacetime.

\section{Conclusions}

In this work, by appealing to  the homothetic and spherical symmetry of a nakedly singular, self-similar Tolman-Bondi spacetime, we have presented a complete analysis of the behaviour of null geodesics on such spacetime. The results of this work confirm the existence of future-directed non-radial null geodesics emanating from the central singularity and are in agreement with those obtained by Mena and Nolan~\cite{bNfM01} and also with our more general results in~\cite{nOoStZ15}. The Hamiltonian treatment employed in this work has proven to be a very efficient technique to obtain many insights into the behaviour of null geodesics. For instance, through this technique, it became clear that the magnitude of the homothetic generator and the area of the two spheres define an effective potential $W(x)$ and a critical impact parameter $\beta_c^2$ which plays a decisive role in the behaviour of radial and non-radial null geodesics as Lemmata~\ref{Lem:W} and~\ref{Lem:Ybetapm} show. Besides the fact that they shed light on the structure and properties of the central singularity of a self-similar Tolman-Bondi spacetime, the results of the present work are important in another respect. They offer the possibility of studying the shadow that a naked singularity casts into the eyes of an asymptotic observer. This analysis would require the interior self-similar collapsing spacetime to be matched to an exterior Schwarzschild vacuum spacetime, a task that can always be accomplished satisfying standard junction conditions (see for instance~\cite{kL00} and references therein). Details of the modeling of the shadow of a naked singularity are discussed elsewhere~\cite{nOoStZ15}.

\ack
O.S. wishes to thank the Perimeter Institute for Theoretical Physics and the Department of Physics at Queen's University, where part of this work was done, for hospitality. T.Z. thanks the Department of Physics at Queen's University and Kayll Lake for the hospitality during a sabbatical year. We are also very grateful to Kayll Lake for many discussion on some of the issues related to this work. 

This research was supported in part by CONACyT Grants No. 232390 and No. 234571, by a CIC Grant to Universidad Michoacana and by Perimeter Institute for Theoretical Physics. Research at Perimeter Institute is supported by the Government of Canada through Industry Canada and by the Province of Ontario through the Ministry of Research and Innovation.

\section*{References}

\bibliographystyle{iopart-num}
\bibliography{../References/refs_collapse}

\providecommand{\newblock}{}
\begin{thebibliography}{10}
\expandafter\ifx\csname url\endcsname\relax
  \def\url#1{{\tt #1}}\fi
\expandafter\ifx\csname urlprefix\endcsname\relax\def\urlprefix{URL }\fi
\providecommand{\eprint}[2][]{\url{#2}}

\bibitem{vV03}
Nieto J, Saucedo J and Villanueva V 2003 {\em Phys. Lett.\/} {\bf A312}
  175--186

\bibitem{vV07}
Nieto J, Saucedo J and Villanueva V 2007 {\em Rev. Mex. Fis.\/} {\bf 53}
  141--145

\bibitem{vV94}
Nieto J and Villanueva V 1994 {\em Nuovo Cim.\/} {\bf B109} 821--827

\bibitem{IPTA}
 http://wwwipta4gworg International pulsar timing array

\bibitem{vV13}
Nieto J, Leon E and Villanueva V 2013 {\em Int. J. Mod. Phys.\/} {\bf D22}
  1350047

\bibitem{dElS79}
Eardley D and Smarr L 1979 {\em Phys. Rev. D\/} {\bf 19} 2239--2259

\bibitem{dC84}
Christodoulou D 1984 {\em Comm. Math. Phys.\/} {\bf 93} 171--195

\bibitem{rN86}
Newman R 1986 {\em Class. Quantum Grav.\/} {\bf 3} 527--539

\bibitem{pJiD93}
Joshi P and Dwivedi I 1993 {\em Phys. Rev. D\/} {\bf 47} 5357--5369

\bibitem{Joshi-Book}
Joshi P 2008 {\em Gravitational Collapse and Spacetime Singularities\/}
  (Cambridge: Cambridge University Press)

\bibitem{bNfM01}
Mena F and Nolan B 2001 {\em Class. Quantum Grav.\/} {\bf 18} 4531--4548

\bibitem{nOoS11}
Ortiz N and Sarbach O 2011 {\em Class. Quantum Grav.\/} {\bf 28} 235001 (27pp)

\bibitem{mCaT71}
Cahill M~E and Taub A~H 1971 {\em Comm. Math. Phys.\/} {\bf 21} 1--40

\bibitem{kLtZ90}
Lake K and Zannias T 1990 {\em Phys. Rev. D\/} {\bf 41}(12) 3866--3868

\bibitem{aOtP87}
Ori A and Piran T 1987 {\em Phys.Rev.Lett.\/} {\bf 59} 2137

\bibitem{dC94}
Christodoulou D 1994 {\em Annals Math.\/} {\bf 140} 607--653

\bibitem{pB95}
Brady P~R 1995 {\em Phys. Rev. D\/} {\bf 51}(8) 4168--4176

\bibitem{bNfM02}
Nolan B and Mena F 2002 {\em Class. Quantum Grav.\/} {\bf 19} 2587--2605

\bibitem{bCaC05}
Carr B and Coley A 2005 {\em General Relativity and Gravitation\/} {\bf 37}
  2165--2188

\bibitem{bNtW02}
Nolan B and Waters T 2002 {\em Phys. Rev. D\/} {\bf 66} 104012

\bibitem{tWbN09}
Nolan B and Waters T 2009 {\em Phys. Rev. D\/} {\bf 79} 1084002

\bibitem{eDbN11}
Duffy E and Nolan B 2011  ArXiv:1108.1103 [gr-qc]

\bibitem{eDbN11b}
Duffy E and Nolan B 2011 {\em Class. Quantum Grav.\/} {\bf 28} 105020 (30pp)

\bibitem{nOoStZ15}
Ortiz N, Sarbach O and Zannias T 2015  In preparation

\bibitem{kL00}
Lake K 2000 {\em Phys. Rev. D\/} {\bf 62} 027301

\end{thebibliography}

\end{document}